\documentclass[%
 amsmath,amssymb,
 twocolumn,showpacs,floatfix,
 aps,prl 
]{revtex4-1}

\usepackage{graphicx}
\usepackage{dcolumn}
\usepackage{bm}
\usepackage{color}

\begin{document}

%
\title{
Control of Competing Superconductivity and Charge Order by Non-equilibrium Currents
}

\author{Anne Matthies}
\author{Jiajun Li}%
\author{Martin Eckstein} 
\affiliation{%
Department of Physics, University of Erlangen-N\"urnberg, 91058 Erlangen, Germany
}%

\date{\today}
\newcommand*{\MyFigurePath}{../MyPlotting}
\begin{abstract}
We study the competing charge-density-wave and superconducting order in the attractive Hubbard model under a voltage bias, using steady-state non-equilibrium dynamical mean-field theory. We show that the charge-density-wave is suppressed in a current-carrying non-equilibrium steady state. This effect is beyond a simple Joule-heating mechanism and a ``supercooled" metallic state is stabilized at a non-equilibrium temperature lower than the equilibrium superconducting $T_c$. On the other hand, a current-carrying superconducting state is dissipation-less and thus not subject to the same non-thermal suppression, and can therefore nucleate out of the supercooled metal, e.g. in a resistive switching experiment. The fact that an electric current can change the relative stability of different phases compared to thermal equilibrium, even when a system appears locally thermal due to electron-eletron scattering, provides a general perspective to control intertwined orders out of equilibrium. 
\end{abstract}

\pacs{71.30.+h,71.45.Lr,74.25.-q}
\maketitle

Strongly correlated materials often have rich phase diagrams resulting from the complex interplay of structural, magnetic and electronic orders.  In many cases, potentially interesting states are suppressed by competing phases which are thermodynamically in close proximity. In particular, competition of superconductivity and charge order is observed in a wide range of systems, including high-$T_{c}$ cuprates \cite{tranquada1995, kivelson2003, chang2012, ghiringhelli2012, gabovich2010}, transition metal dichalcogenides \cite{Berthier1976, Gabovich2001, Sipos2008, Bawden2016}, or oxides interfaces  \cite{Frano2016}. How to control and distinguish such competing phases arises as an interesting question which is currently under intense research.

Nonequilibrium excitations, including intense laser or current pulses, provide an intriguing pathway to reveal ``hidden'' states which are inaccessible under equilibrium conditions. For example, a suppression of  charge order may underly the recent observation of light-induced superconductivity \cite{Fausti2011,Nicoletti2014}, and hidden charge density wave states have been prepared with strong current pulses \cite{Vaskivskyi2016}. The identification of generic pathways of non-equilibrium control is however challenging: Excited electron distributions in correlated systems often rapidly evolve towards a quasi-thermal ``hot electron'' state (sometimes within femtoseconds \cite{Ligges2017}), but merely increasing an effective electronic temperature above the melting temperature of a dominant phase can obviously not reveal a subdominant order with a lower transition temperature. Although tantalizing results for the dynamical interplay of multiple orders have been obtained for the pre-thermal electron dynamics, where collisionless mean-field descriptions can be used \cite{Fu2014, Dzero2015, Krull2016, Sentef2017, Murakami2017}, the understanding of a robust non-equilibrium mechanism for the suppression of competing phases on times longer than the electronic thermalization time remains an open question.

In this work, we focus on a non-equilibrium steady state (NESS), i.e., the {\em long-time behavior} reached by a system which is simultaneously subject to an electric field and coupled to a dissipative environment. In a model where charge density wave (CDW) and superconducting phases have the same critical temperature T$_c$ in equilibrium, we demonstrate that the non-equilibrium electric current $J$ can act as a control parameter to suppress the pure CDW phase. In spite of strong electron-electron scattering, which causes the electronic state to be locally close to thermal, the mechanism is different from heating, and the resulting normal phase has a temperature below T$_c$. The superconductor, in contrast, can carry the same current $J$ without dissipation. An NESS can be realized in resistive switching experiments, as discussed below, but due to rapid electronic thermalization also pulses of nanosecond or even picosecond duration may be described along these lines.
 
Manipulating strongly correlated materials through electric fields has a long and successful history. One of the mostly explored phenomena in this context is the voltage-driven insulator-metal transition, which exists widely in transition metal oxides/dichalcogenides and other correlated insulators \cite{janod2015}. The thermal scenario, in which Joule-heating increases the non-equilibrium temperature and causes the transition is discussed \cite{zimmers2013, chudnovskii1998, duchene1971,li2015}, but also non-Fermi-Dirac distributions of hot electrons can play a critical role \cite{li2017}. In the present work, we use the steady-state formulation of non-equilibrium dynamical mean-field theory (DMFT) \cite{aoki2014,amaricci2012,li2015} to capture both correlation effects and strong electron-electron scattering.

{\em Model --} As a minimal model with competing superconducting and CDW phases we consider the half-filled attractive Hubbard model on a bipartite lattice with two sub-lattices $A$ and $B$. The Hamiltonian is given  by
\begin{align}
H
=  -t_0
\sum_{\langle i,j\rangle,\sigma}c^{\dagger}_{i\sigma} c^{\phantom \dagger}_{j\sigma} 
+
U
\sum_{j}
(n_{j\uparrow}-\tfrac12 )(n_{j\downarrow}-\tfrac12),
\label{HHM}   
\end{align}
where $c_{j\sigma}^\dagger$ creates an electron with spin $\sigma$ on lattice site $j$, $n_{j\sigma}=c_{j\sigma}^\dagger c_{j\sigma}$ and $n_j=n_{j\uparrow}+n_{j\downarrow}$; $t_0$ is the hopping between nearest neighbour sites, and $U<0$ is an attractive on-site 
interaction. In equilibrium, the model is characterized by order parameters $\psi_{\rm CDW}=\langle n_j\rangle_{j\in A}-\langle n_j\rangle_{j\in B}$ for the CDW and  $\psi_{\rm SC}(j)=\langle c_{j\uparrow}c_{j\downarrow}\rangle$ for superconductivity, respectively.  At half-filling ($n=1$) the two orders are degenerate \cite{micnas1990, freericks1993, toschi2005}. Throughout the paper we will focus on this regime. To simulate the NESS in the normal metal and in the CDW phase, the system is subject to a bias voltage and coupled to a heat bath, which is taken to be a Fermion reservoir with flat density of states $\Gamma$ and fixed temperature $T_b=1/\beta$ (see below) \cite{tsuji2009}. The bias acts as a term $\sum_{j} V(z_j) n_{j}$ in Eq.~\eqref{HHM}, where each site $j$ lies on a layer $z_j\in\{0,\pm1,\pm2,..\}$ of the lattice, and $V\equiv V(z+1)-V(z)$ is the voltage difference between adjacent layers. The system is infinitely extended, so that boundary effects due to the leads can be ignored. A current-carrying state in the superconductor, in contrast, is an equilibrium state with nonzero phase twist $\phi$ between the layers, i.e., $\psi_{\rm SC}(j)= |\psi_{\rm SC}| e^{i\phi z_j}$. 

{\em DMFT Setup --} To study the time-translationally invariant NESS, we use the steady-state formulation of  non-equilibrium DMFT in terms of Keldysh Green's functions $\hat G_{ij}(\omega)$. (For a detailed introduction, see Ref.~\cite{aoki2014}).  In DMFT, the self-energy  $\hat \Sigma_{ij}(\omega)=\delta_{ij}\hat \Sigma_j(\omega)$ is taken to be local, and the Green's function at a given site $j$ in the layered structure is thus given by
\begin{align}
\hat G_j^{-1}(\omega) = \omega- \epsilon_j - \hat \Sigma_j(\omega) - \hat \Gamma(\omega) - 
\sum_{\pm} \hat F^\pm_j(\omega).
\end{align}
Here $\epsilon_j=\frac{1}{2}U(\langle n_{j}\rangle-1)$ is the on-site Hartree energy, and $\hat \Gamma(\omega)$ is the heat bath, given by a constant spectral function $\Gamma^R(\omega)=-i\Gamma$, and $\Gamma^K(\omega)=-2i\Gamma \tanh(\beta\omega/2)$ \cite{tsuji2009}.  The terms $\hat F_j^\pm(\omega)$ describe the hybridization of site $j$ with the neighboring layer with larger ($+$) and lower ($-$) potential, respectively. We take $\hat F^\pm_j(\omega)=t_0^2 G_{j'}(\omega)$ where $j'$ is a neighbor of $j$ on the layer with $z_{j'}=z_{j}\pm 1$, respectively. This corresponds to embedding the layered structure into an infinitely-coordinated Bethe lattice with half-bandwidth $W=2\sqrt{2}t_0$, such that, similar to a cubic lattice with potential gradient along the (111) direction, each site has only neighbors in positive and negative field direction \cite{suppl}. (The hopping $t_0=1$ sets the energy scale.) We also performed analogous simulations for a one-dimensional chain and obtained qualitatively identical results. The DMFT equations are therefore closed by $F^{\pm}_{A(B)}(\omega)=t_0^2 G_{B(A)}(\omega \pm V)$, where we have taken into account that neighboring layers are shifted in energy by the bias $V$ and belong to opposite sub-lattices $A,B$. The local self-energy $\hat \Sigma_j$  is given by the solution of an impurity problem with hybridization function $\hat \Delta_j(\omega)=\hat \Gamma_j(\omega) + \hat F^+_j(\omega) +  \hat F^-_j(\omega)$. We use second order perturbation theory (IPT) to obtain the impurity self-energy. While IPT is not a conserving approximation, we focus on interactions $|U|$ which are small enough such that unphysical behavior as observed in transient simulations at large $U$ \cite{Eckstein2010b} does not occur. From the DMFT solution, we obtain the order parameter  $\psi_{\rm CDW} = i\int \frac{d\omega}{2\pi} [G_A^<(\omega)-  G_B^<(\omega)]$ and the current density $J=2\text{Im} \int \frac{d\omega}{2\pi}  (\hat G_A(\omega)\hat F^+_A(\omega))^<=2\text{Im} \int \frac{d\omega}{2\pi}  (\hat G_B(\omega) \hat F^-_B(\omega))^<$.  The study of superconductivity is performed in the AB-symmetric phase with no bias voltage. In this case, the Green's functions acquire a $2\times2$ Nambu structure in addition to the Keldysh indices. When we assume a phase twist of the order parameter between different layers, the self-consistency reads $\Delta(\omega)=t_0^2 \sum_{\pm} e^{\pm i\phi \sigma_z} G(\omega)e^{\mp i\phi \sigma_z} $, where $\sigma_z$ is the Pauli matrix in Nambu space.


\begin{figure}
	\includegraphics[width=\columnwidth]{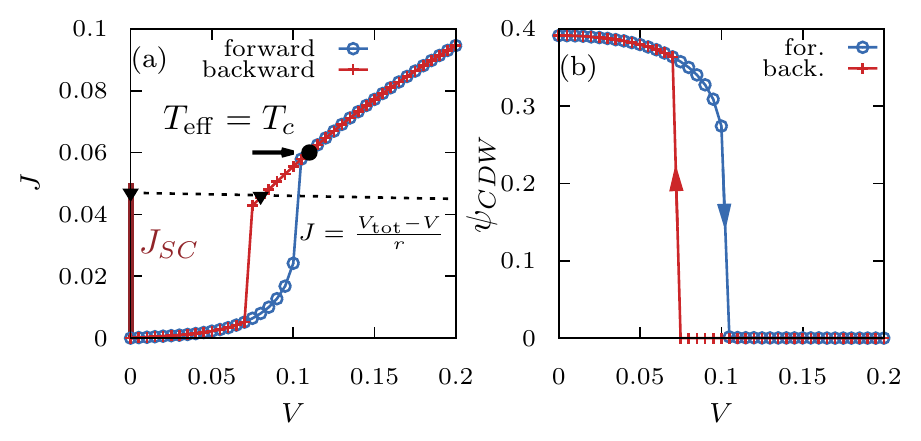}
	\caption{\label{fig:IV-curves} 
{\bf (a)} Current $J$ as function of bias $V$, for $U=-2.5$, bath temperature $T_{b}=0.01$, and $\Gamma=0.1$. The ``forward" (``backward") curves are found by continuously following the DMFT solution upon increasing (decreasing) the field. The red vertical line shows the supercurrent $J_{\rm SC}$ in equilibrium. The black dashed line illustrates the linear relation between $J$ and $V$ when the sample is connected to an external resistor $r$ and a total bias $V_\text{tot}$ is applied, and  the inverted triangles label the two possible physical solutions  (see text).  The black arrow indicates the state in which the metal has an effective temperature $T_c$. {\bf (b)} CDW order parameter $\psi_{\rm CDW}$ for the same setting as (a).}
\end{figure}

\begin{figure}
\includegraphics[width=\columnwidth]{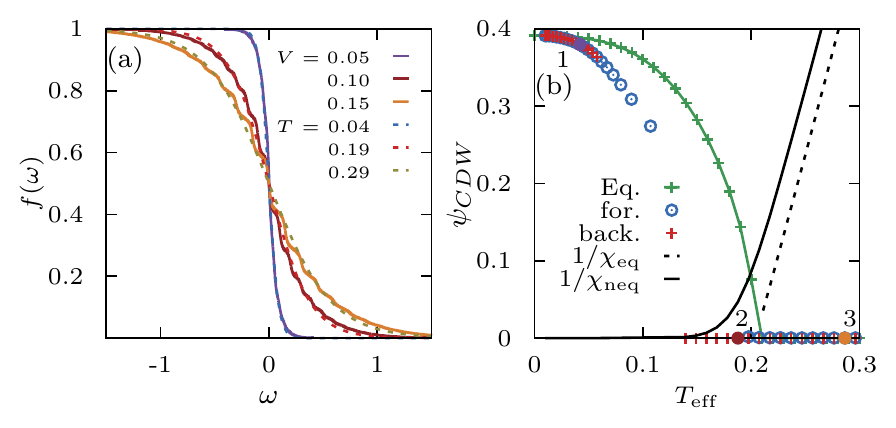}
\caption{\label{fig:density} 
{\bf (a)} Distribution functions for $V_1=0.05$ (CDW), $V_2=0.1$ (coexistence regime) and $V_3=0.15$ (metal), and Fermi-Dirac distributions function for indicated temperatures. ($U=-2.5$,  $T_b =0.01$, $\Gamma=0.1$, as in Fig.~\ref{fig:IV-curves}) {\bf (b)} Order parameter $\psi_{\rm CDW}$ in the forward/backward NESS and in equilibrium, plotted against effective temperature $T_{\rm eff}$ and equilibrium temperature, respectively;  $1/\chi_{eq}$ and  $1/\chi_{neq}$ show the CDW susceptibilities in the equilibrium and backward metallic state (scaled by the same arbitrary unit). Circle symbols 1-3 indicate the parameters of the distribution functions shown in (a).}
\end{figure}

\emph{Results -- }
To analyse phases with a given current $J\neq0$, we must distinguish the case of a dissipative current ($V\neq 0$) and a non-dissipative current ($V=0$). We first present an analysis of the current-induced melting of the CDW at $V\neq0$. As superconducting transport is not dissipative in the bulk (and an AC Josephson effect between the layers is excluded), we can  restrict this  analysis to the pure CDW and normal phases. We first exemplarily discuss results for on-site interaction $U=-2.5$, bath temperature $T_b=0.01$, and damping $\Gamma=0.1$. In Fig.~\ref{fig:IV-curves} we show the current (a)  and the corresponding order parameter $\psi_{\rm CDW}$ (b) in the NESS as a function of the bias voltage. The system remains insulating when $V$ is not large enough to overcome the gap (there is a small current due to Zener tunnelling across the gap). With increasing $V$, the CDW order becomes unstable and the system finally transits into a metallic state. In a certain range of bias voltages, both metallic and CDW phases can exist. This indicates that the transition is of first order, in contrast to the equilibrium transition (c.f. Fig.~\ref{fig:density}b for $\psi_{CDW}$ in equilibrium). The coexistence can be explored by a hysteresis: By increasing $V$ in steps, using the previous solution as a seed for the DMFT iteration, the system undergoes an insulator-to-metal transition (IMT) at an upper critical voltage $V_{u}$ (blue circles). Decreasing $V$ drives the system to a metal-to-insulator transition (MIT) at a lower critical voltage $V_{c}$ (red crosses). 

We note that the two transitions arise from distinct mechanisms \cite{sugimoto2008,li2017,han2018}. The IMT is induced when charge excitations created by the external field break the ordered phase, whereas the MIT occurs when the metallic state becomes unstable to infinitesimal CDW fluctuations. Hence one can expect that the CDW susceptibility diverges as $V$ approaches the lower critical $V_{c}$ from above, in line with the second-order CDW-metal transition in equilibrium. Furthermore, in a potential thermal scenario one would expect that the metal can be described well by an effective temperature $T_{\rm eff}$ at the {\em lower} critical $V_{c}$ which is close to the equilibrium critical temperature $T_c$ \cite{sugimoto2008,li2017}. To test this scenario and clarify the mechanism of the field-driven transition, we analyze the electronic distribution functions $f(\omega)=-\frac{1}{2}\operatorname{Im} G^<(\omega)/\operatorname{Im} G^r(\omega)$ throughout the transition regime. Figure~\ref{fig:density}a displays $f(\omega)$ for three representative points on the $J-V$ curve: in the CDW phase ($V_1=0.05$), the coexistence region ($V_2=0.1$) and the metallic phase ($V_3=0.15$). The dashed lines show a fit with a Fermi-Dirac distribution. The non-thermal distribution functions feature a slight step-like deviation from the Fermi-Dirac distribution (in particular at larger bias $V$), but nevertheless the thermal distributions provide a reasonable fit, which can be explained by the strong electron-electron scattering. For the further analysis we therefore use the Fermi Dirac fit to define an effective temperature $T_{\rm eff}$ of the NESS.

To analyze whether the melting of the CDW order is dominated by the thermal mechanism, we plot in Fig.~\ref{fig:density}b the order parameter as a function of the effective temperature. As the switching at $V=V_{c}$ occurs, $T_{\rm eff}$ is indeed of similar order of magnitude as compared to the  equilibrium $T_c$. However, the non-thermal state is clearly beyond the solely thermal description: 
The CDW order in the NESS with effective temperature $T_{\rm eff}$ is consistently weaker than the equilibrium order at $T=T_{\rm eff}$. Moreover, $T_{\rm eff}$ at $V=V_{c}$ is lower than $T_c$, resulting in a ``supercooled'' metallic state below $T_c$. Finally, we can compare the charge-density-wave susceptibilities $\chi_{neq}$ and $\chi_{eq}$, which are obtained from the ratio $\chi=\psi_{\rm CDW}/\varepsilon_{\rm CDW}$ after computing $\psi_{\rm CDW}$ in the presence of a small staggered field $\varepsilon_{\rm CDW}$ in the NESS and in equilibrium, respectively. The non-equilibrium $\chi_{neq}$ is finite throughout the supercooled phase (indicating stability of this phase), but $\chi_{neq}$ in a NESS with $T_{\rm eff}$ is always lower than $\chi_{eq}$ at temperature $T=T_{\rm eff}$ (Fig.~\ref{fig:density}b).

\begin{figure}
		\includegraphics[width=\columnwidth]{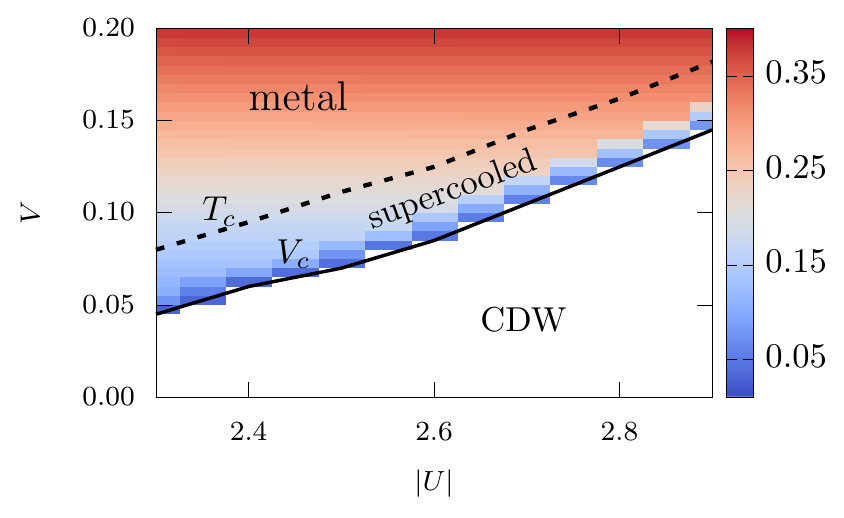}
\caption{\label{fig:effT} 
Effective temperature map for different $U$ and $V$. The solid line indicates the (lower) critical bias $V_c$, whereas the dashed line shows the critical temperature $T_c$. Below the line $V_c$ the metallic phase becomes unstable against CDW formation. There exists a region where the metal has temperature $T_{\rm eff} <T_c$ for a wide range of Hubbard $U$. }
\end{figure}

The reduction of the order parameter and the susceptibility in the NESS compared to an equilibrium state at the same effective temperature show that the non-equilibrium current can control the suppression of the CDW independent of a thermal mechanism. The result can be summarized in a nonequilibrium phase diagram. In Fig.~\ref{fig:effT} we show a false color map of the effective temperature as function of different interaction strengths and voltages. The CDW and metallic phases are separated by the critical voltage $V_c$ (solid line). In the metallic phase, the effective temperature increases monotonically with increasing bias, and the supercooled regime is enclosed by the two lines $V_c$ and $T_{\rm eff}=T_c$ (dashed line). The corresponding upper bound $V(T_c)$ for the supercooled states is systematically higher than the critical voltage $V_c$ for the MIT switching, with almost constant $V(T_c)-V_c\approx0.04$ in the considered $U$ regime. We thus observe a robust supercooling effect for a wide range of parameters at weak-coupling. 

\begin{figure}
	\includegraphics[width=\columnwidth]{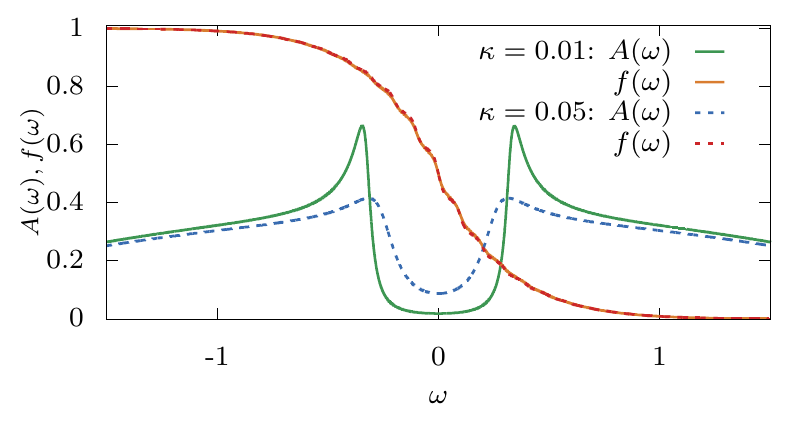}
	\caption{\label{fig:SpecFunc} 
Spectral function $A(\omega)$ of a superconducting state at $U=-2.5$, coupled to a supercooled bath, whose distribution function is taken from the NESS at $V=0.1$, see Fig.~\ref{fig:density}a. ($\kappa$ is the coupling strength of the bath). The spectral function shows a superconducting gap, the order parameter is $\psi_{\rm SC}=0.141,0.182$ for $\kappa=0.05,0.01$, respectively. The distribution function of the superconducting state is also shown for comparison.
}
\end{figure}

A current-controlled resistive switching experiment provides a promising way to explore the different nonequilibrium phases. In this setup, an external resistor $r$ is connected to the sample, and the total voltage $V_\text{tot}=V+rJ$ is controlled \cite{janod2015}. While $V_\text{tot}$ is adjusted, the intersection of $J=(V_\text{tot}-V)/r$ and the $J-V$ curve determines the physical solutions, as illustrated in Fig.~\ref{fig:IV-curves}. Ideal voltage-controlled and current-controlled experiments correspond to the two limits $r=0$ and $\infty$, respectively. Focusing on the non-superconducting phases first, by increasing $V_\text{tot}$ from zero, the dashed line in Fig.~\ref{fig:IV-curves} would shift upward from zero and the operating point (marked as inverted triangle in Fig.~\ref{fig:IV-curves}) would move along the insulating $J-V$ branch, as the case of normal field-driven insulator-metal transition \cite{ridley1963}. Above a first critical $V_\text{tot}$, no intersection with either the metallic or insulating solution is possible, so that the system must form a filamentary (phase separated) conducting phase \cite{duchene1971,zimmers2013,li2017}. Finally, when $V_\text{tot}$ is further increased, the system resides in the homogeneous supercooled metallic regime at $T_\text{\rm eff}<T_c$. (Depending on the parameters, also a direct jump from the CDW to the metallic phase may occur.)

The superconducting solution can dramatically modify this picture. In contrast to the CDW, a current-carrying superconducting phase is dissipation-less and is not suppressed by the same non-thermal effect. In a current-controlled resistive switching experiment, the superconducting phase is therefore always a possible solution, as long as the current is smaller than the maximal supercurrent $J_{SC}$, which is shown for the present parameters in Fig.~\ref{fig:IV-curves} by the vertical red line at sample bias $V=0$. After the CDW phase is melted and the system is transferred to the metallic state, we can expect that the superconducting phase would nucleate if the metal has an effective temperature $T_{\rm eff} <T_c$. To demonstrate the possibility of this nucleation (in spite of the fact that the distribution function $f(\omega)$ in the supercooled phase shows a slight deviation from a Fermi-Dirac form), we couple the lattice to a bath with the non-thermal distribution function $f(\omega)$ of the supercooled state and the density of states of the metal, and study  the possible superconducting order. In Fig.~\ref{fig:SpecFunc} we display the resulting local spectral function $A(\omega)=-\frac{1}{\pi}\operatorname{Im} G^r(\omega)$,  which shows a clear superconducting gap. We have taken $f(\omega)$ from the metallic state at $V=0.1$, which is close to the upper bound of the supercooled regime. For smaller $V$ superconductivity is quantitatively more robust. (The coupling $\kappa$ of the bath should be kept small, so that the bath just imposes the distribution function. Too large bath couplings suppress superconductivity due to a metallic proximity effect, which can be viewed as an artefact of the Fermion bath taken here.)

\emph{Conclusion --} Using dynamical mean-field theory, we have studied the attractive Hubbard model under a strong voltage bias. We find that a pure CDW is suppressed by the electric current beyond a mere Joule-heating mechanism, and a normal metallic phase can be stabilized against CDW formation even below the equilibrium critical temperature. This current-induced supercooled state may be realized in resistive switching experiments. Since its temperature is also below the superconducting $T_c$, in a current-controlled experiment it can be transferred into a superconducting phase, which can carry the same (but dissipation-less) current. In the half-filled attractive Hubbard model, where CDW and superconducting phases are degenerate in equilibrium, there may also be supersolid phases with the same super-current, which we have not investigated here. While the latter is particular to the Hubbard model, the current-controlled mechanism for the non-thermal suppression of CDW works even against a stabilizing CDW field, which hints that it is also active in a wider range of models where CDW is the more stable phase in equilibrium. This question should be investigated in more realistic models for the materials mentioned in the introduction, including, for example, phonon-mediated interactions instead of the static attractive Hubbard $U$ \cite{schuett2018}, and possibly using more sophisticated impurity solvers \cite{arrigoni2013}. Further studies should also address whether the current-controlled mechanism can be applied to manipulate more general phases involving intertwined charge, orbital, and spin orders \cite{tokura2000,cruz2008,Vaskivskyi2016}. Also it is intriguing to study the intermediate to strong coupling regime, where a bad superconducting phase characterized by a decreased order parameter and incoherent states can emerge under finite current \cite{amaricci2018}. In general, the competition of a `bad' SC phase, in particular with dissipative currents, and the CDW phase in the current-driven regime may
give rise to novel physics beyond the scope of the current
work. To seriously study this possibility, one needs to
treat the two phases on equal footing, for example, in a
finite-size system with coexisting CDW/SC phases. Finally, it will be interesting to investigate the possible switching in real-time by simulating the non-equilibrium dynamics on the time scale of thermalization \cite{aoki2014}.

\begin{acknowledgments}
\emph{Acknowledgments -- }We acknowledge the financial support from the ERC starting grant No. 716648.
\end{acknowledgments}

\bibliography{references.bib}
\end{document}